\documentclass[10pt,journal]{IEEEtran}

\usepackage{url,comment,cite}
\usepackage{amsmath,amssymb,amsfonts,bbm,dsfont}
\usepackage{graphicx}
\usepackage{epsfig,epsf,psfrag,textcomp,tabularx}
\usepackage[usenames,dvipsnames]{xcolor}
\usepackage{subfigure}
\usepackage{float}
\usepackage{stfloats,mathtools}
\usepackage[algoruled,medskip,dontprintsemicolon,linesnumbered,Algorithm]{algorithm}
\usepackage[noend]{algpseudocode}
\usepackage{autonum}
\usepackage{multirow}

\newcommand{\Bc}{\mathcal{B}}
\newcommand{\Ec}{\mathcal{E}}
\newcommand{\Gc}{\mathcal{G}}

\newcommand{\Ic}{\mathcal{I}}

\newcommand{\Qc}{\mathcal{Q}}

\newcommand{\Sc}{\mathcal{S}}
\newcommand{\Zc}{\mathcal{Z}}
\newcommand{\Fc}{\mathcal{F}}

\newcommand{\etav}{\boldsymbol{\eta}}

\newcommand{\muv}{\boldsymbol{\mu }}

\newcommand{\bv}{{\bf b}}

\newcommand{\sv}{{\bf s}}
\newcommand{\tv}{{\bf t}}
\newcommand{\uv}{{\bf u}}
\newcommand{\wv}{{\bf w}}
\newcommand{\vv}{{\bf v}}

\newcommand{\yv}{{\bf y}}

\newcommand{\onev}{{\bf 1}}

\newcommand{\Bm}{{\bf B}}

\newcommand{\Hm}{{\bf H}}
\newcommand{\Id}{{\bf I}}

\newcommand{\EE}{\mathbb{E}} %

\def\Herm{^\mathsf{H}}

\newcommand{\ee}{{\rm e}}
\newcommand{\jj}{{\rm j}}  %
\def\non{\nonumber\\}

\newcommand{\Fig}[1]{Fig.~\ref{fig:#1}}
\newcommand{\Sec}[1]{Sec.~\ref{sec:#1}}
\newcommand{\Tab}[1]{Tab.~\ref{tab:#1}}
\newcommand{\Eq}[1]{(\ref{eq:#1})}

\newtheorem{lemma}{Lemma}

\begin{document}

\title{A Belief Propagation Solution for Beam Coordination in MmWave Vehicular Networks} 

\author{Zana~Limani~Fazliu,~\IEEEmembership{Member,~IEEE,}
Francesco~Malandrino,~\IEEEmembership{Senior Member,~IEEE,}
Carla~Fabiana~Chiasserini,~\IEEEmembership{Fellow,~IEEE,} 
 Alessandro~Nordio,~\IEEEmembership{Member,~IEEE} %
\thanks{Z.~Limani~Fazliu is with  University of Prishtina, Prishtina, Kosovo.
F.~Malandrino, C.~F.~Chiasserini and A.~Nordio are with CNR-IEIIT, Italy.
F.~Malandrino and C.~F.~Chiasserini are with CNIT, Italy.
C.~F. Chiasserini is with  Politecnico di Torino.

This work was supported by the EU Commission under the RAINBOW project
(Grant Agreement no.\,871403).
The views expressed are those of the authors and
do not necessarily represent the project. The Commission is not liable
for any use that may be made of any of the information contained
therein.

}
}

\maketitle

\begin{abstract}
Millimeter-wave communication is widely seen as a promising option to increase the capacity of vehicular networks,
where it is expected that connected cars will soon need to transmit and receive large amounts of data.
Due to harsh propagation conditions, mmWave systems resort to narrow beams to serve their users, and such beams need to be configured according to traffic demand and its spatial distribution, as well as interference. 
In this work, we address the beam management problem, considering an urban vehicular network composed of gNBs. We first build an accurate, yet tractable, system model and formulate an optimization problem aiming at maximizing the total network data rate while accounting for the stochastic nature of the network scenario. 
Then we develop a graph-based model capturing the main system characteristics and use it to develop 
a belief propagation algorithmic framework, called CRAB, that 
has low complexity and, hence, can effectively cope with large-scale scenarios.   
We assess the performance of our approach under real-world settings and show that, 
in comparison to state-of-the-art alternatives, CRAB provides on average a 50\% improvement 
in the amount of data transferred by the single gNBs and up to 30\% better user coverage.
\end{abstract}


\section{Introduction}
\label{sec:intro}
Vehicular networks and their users have long been identified as great consumers of data, 
for applications including safety~\cite{bloessl2015scrambler,sommer2014shadowing}, 
map updates~\cite{leontiadis2010extending}, content downloading~\cite{malandrino2012optimal}, 
and onboard entertainment~\cite{yu2013toward}. The issue has been further exacerbated by 
the emergence of connected and autonomous vehicles: such vehicles need frequently-updated 
and detailed information on the topology and conditions of the 
road~\cite{uhlemann2015autonomous,yu2016connected}, in addition to providing their users 
with even richer multimedia content, especially for automated and autonomous 
vehicles~\cite{uhlemann2015autonomous}. The effect of such trends is a further increase 
of the requirement posed on the infrastructure serving the vehicles.

Whenever more network capacity for wireless networks is needed, moving towards higher 
frequencies is an appealing option. Indeed, millimeter-wave (mmWave) networks, which 
operate at frequencies of tens of gigahertz and were originally envisioned for quasi-static, 
indoor scenarios, are becoming  an appealing option also for vehicular networks. However, although 
mmWave technology 
allows for very large bandwidths and high data rate, it is also characterized by 
harsh propagation conditions, with high path loss and virtually no connectivity in 
non-line-of-sight conditions.

To address these shortcomings, directional antennas capable of {\em beamforming} are employed. 
Unlike antennas used at lower frequencies, mmWave base stations (gNBs) serve 
their users through {\em beams}, each concentrating the available power along a given direction 
in order to achieve higher values of 
received power as well as lower interference~\cite{roh2014millimeter}. This also means that 
swift, high-quality beamforming decisions are critical to the performance, and indeed the 
very usefulness, of mmWave networks. The task of beamforming is especially challenging in 
vehicular networks, owing to the fast mobility of the users; on the positive side, such 
mobility is constrained by the road topology and can be forecast with good 
accuracy~\cite{ketabi2019vehicular}.

In this work, we formulate the problem of beamforming in vehicular networks as an optimization 
problem, where the decisions to make concern the beam configuration at each gNB, and the objective 
is to maximize the total network data rate. Owing to the multiple sources of variability in our 
scenario and the stochastic nature of the wireless medium, we then introduce a {\em randomized} 
solution strategy. Under such an approach, we do not choose directly a beam configuration, but 
rather (i) we set the {\em probabilities} of each configuration to be selected, and (ii) we 
enact the beam configurations over time  according to those probabilities. Similar randomized 
approaches have been successfully used in many fields, including routing~\cite{mukhopadhyay2015power}, 
resource provision in cloud computing~\cite{zhang2014dynamic}, and network 
orchestration~\cite{zhou2019online}. The intuition behind the success of randomized approaches 
is that, in very complex scenarios, different sources of randomness tend to cancel one another, 
in a manner similar to errors in Fermi approximation. Therefore, randomly choosing the beam 
configuration to enact can actually reduce the negative effects of the variability in vehicular 
traffic or wireless channel conditions.

We further introduce a distributed heuristic algorithm, called coverage-rate aware belief propagation (CRAB), 
to make effective and efficient decisions about the beam configuration probabilities. CRAB is 
based upon the belief propagation approach~\cite{yedidia2003understanding,ihlerloopy}, and 
leverages scenario-specific information and insights. 
In CRAB, gNBs exchange {\em messages} about their {\em local} beam 
configurations, until converging to a situation where the {\em global} data rate is maximized. 
We compare CRAB against state-of-the art approaches based upon clustering or graph matching, and 
find it to provide remarkably better solutions with a very low complexity.

In summary, our main contributions are as follows:
\begin{itemize}
\item We provide a detailed, yet mathematically tractable, model of a mmWave network infrastructure for 
the support of vehicular communications, which is based upon established standards and cutting-edge 
research studies. Using this model, we formulate both a centralized and distributed 
optimization problem making beamforming decisions that maximize the overall network data rate. 
Unlike previous work and motivated by the highly dynamic scenario under study, 
we leverage a randomized approach 
whereby decisions concern the probability with which a given configuration is used at each gNB. 

\item We then focus on the distributed formulation, so as to make high-quality 
decisions while exploiting solely local information. In so doing, we develop a graph-based representation
of the network infrastructure, which accounts for the inter-gNB conflicts, i.e., interference and overlapping user coverage.
Through such a model, we define a novel belief propagation-based approach, named CRAB, that efficiently yields 
beamforming configurations at the gNBs that effectively avoid inter-gNB conflicts.

\item We assess the performance of the CRAB scheme and compare it against state-of-the art alternatives, 
under real-world settings. Our results show that under CRAB  over 30\% of the gNBs experience 
a rate increase of over 100\%, and 55\%  experience a gain of at least 50\%, 
 while serving up to 30\% more vehicular users and exhibiting higher fairness. 
\end{itemize}

The remainder of this paper is organized as follows. We begin by discussing some relevant
related work in \Sec{rel-work}, highlighting the novelty of our approach, and, then, we introduce 
our system model in \Sec{system}. In \Sec{problem-formulation}, we formulate a centralized and distributed 
beamforming optimization problem. 
\Sec{solution} presents the CRAB algorithm, highlighting how it combines the mechanics 
of message-passing algorithms with scenario-specific knowledge and insights. 
The performance of CRAB is compared against that of state-of-the-art alternatives 
in \Sec{results}, while \Sec{conclusion} concludes the paper.

\section{Related Work\label{sec:rel-work}}

Beam management in mmWave networks, and in particular initial access, beam alignment and configuration have been active areas of research in the last few years. While earlier works focused on exhaustive and iterative search techniques to identify and configure mmWave beam directions \cite{ia-wei, ia-giordani}, later works turned to more intricate approaches which were often driven by data and based on advanced learning techniques \cite{deepbeam-polese, lidar-zecchin, learnandadapt-hussain, lightweightprl-vanhuynh, ponnada2021}. In particular, 
\cite{deepbeam-polese} adopts a data-driven approach and uses convolutional neural networks to reduce 
the necessary coordination between transmitter and receiver when configuring their beam settings. 
A similar data-driven approach is used in \cite{lidar-zecchin} wherein
a convolutional neural network architecture is applied in tandem with 
LIDAR preprocessing technique to optimize beam selection. The proposed model is trained to exploit LIDAR
and positional data in order to identify the best beam directions and reduce the beam search overhead in 
vehicle-to-infrastructure  communication.

In \cite{learnandadapt-hussain}, the authors tackle the optimal beam selection problem by formulating 
the decision-making process as a partially observable Markov decision
process. They also propose a point-based value iteration
 method to design an approximately optimal policy, wherein the goal is to select the strongest 
 beam pair that maximizes the beamforming gain between a single base station-user pair. 
The study in \cite{lightweightprl-vanhuynh} envisions an optimal beam association policy 
in mmWave vehicular networks using a lightweight alternative to the Q-learning algorithm, 
while modeling the dynamics of the mmWave communication link using a semi-Markov decision process framework. 
\cite{lightweightprl-vanhuynh}, however, deals only with  straight-road scenarios, with the assumptions 
that infrastructure nodes cover separate and distinct segments of the road. 
The algorithm is therefore independently applied by each node to identify the optimal beam association 
strategy for vehicles under their coverage. No coordination between nodes is foreseen.

To predict the best beam choice for a vehicle,  
 \cite{ponnada2021}  introduces the usage of channel charting.  
 The proposed approach consists of two stages: one offline during which the channel charts are 
 constructed for each beam, and one online, during which live collected data is used to make online 
 predictions for the best beam combination. Again, the work in  \cite{ponnada2021} focuses on a single infrastructure node 
 covering a single straight road segment.

Learning-based techniques, however, are known to be computationally taxing and time consuming, 
which is why most of these works consider limited scenarios with a single infrastructure node or  
address the best beam selection with respect to a single vehicle. In addition, they address  
 a highway or a straight road scenario, thereby largely ignoring the interactions and interference 
 potential that can be found in an urban setting.  

Furthermore, all of the above  studies focus
on beam aligning for a single base station-vehicle pair, implicitly assuming
that each mmWave beam is employed to transmit to a single
user only. However, in ultra dense scenarios, a narrow beam can cover several users
simultaneously, users that can be multiplexed within the same
beam. There are only few works that consider that {\em a mmWave link can be used to establish communication 
with several end users simultaneously} \cite{noi-isj20, adptbmfrming-zhang}. 
In particular, \cite{noi-isj20} considers a dense urban scenario, and uses traffic light information 
to guide the beam directions chosen by the infrastructure nodes; however, \cite{noi-isj20} does not 
consider coordination between nodes as we do in this work.  In \cite{adptbmfrming-zhang}, instead, 
the authors focus on vehicle-to-vehicle networks and propose an adaptive beamforming scheme 
based on K-means clustering for point-to-multipoint communications for message dissemination. 
The work in \cite{adptbmfrming-zhang}, however, is tailored to highway scenarios and 
 message dissemination therein is enabled by data relaying performed by individual vehicles, and it 
 cannot be easily extended to vehicle-to-infrastructure communication scenarios.

We also remark that, to our knowledge, few works  have applied graph theory
to address beam management in mmWave communications \cite{graph-ziyuan, mmwave-graph-icc, noi-mobihoc20,mp-fan}. 
Both \cite{graph-ziyuan} and  \cite{mmwave-graph-icc}  apply graph techniques to reduce inter-cell
interference, which is different from  our  preliminary work in \cite{noi-mobihoc20} and this work.  
We recall that, in \cite{noi-mobihoc20} as well as this work, the goal of the proposed graph-based approaches 
is to maximize the network data rate, 
the difference being that in \cite{noi-mobihoc20} a centralized approach is proposed, 
while in this work we consider a distributed approach enabled through the coordination between 
connected nodes in the graph.

As for message passing applied to mmWave networks, studies that leverage such an approach can be found in \cite{mp-fan, mp-myers}. 
In \cite{mp-fan}, the authors use a graph approach to tackle the user association and power control 
in mmwave HetNets, by modeling the network as a coordination graph with edges between base stations and users. 
Using this graph, they apply a message passing algorithm combined with reinforcement learning to 
achieve a solution that maximizes  the overall time averaged risk averse rate of the network. 
The authors do not address the management of the beams in such a network, rather they consider the 
beam pattern of the base station to be fixed and, consequently, users transition from an aligned to a 
non-aligned state as they move. 
The work in \cite{mp-myers}, instead,  uses a dynamic compressed sensing-approximate message passing   
algorithm, to leverage the sparsity and correlation in subchannels
for channel estimation and propose an alternative technique that exploits information about the antenna
geometry and the range of the transceiver distance, for compressive beam alignment. 
The message passing algorithm is used to establish an individual short-range link between one access 
point and its users, and, thus, it does not take into account the behavior of other transmitters 
in the network.

{\em Novelty.} 
Our work represents an improvement over existing literature along  
three main directions. First, our graph-based representation of the mmWave infrastructure 
is a complete and compact way to account for the non-trivial outcome of beamforming decisions, 
without the intrinsic complexity of data-driven approaches. Second, our distributed message-passing 
solution strategy allows for swifter convergence compared to centralized algorithms, 
without the need to share and transfer large amounts of data. Third, by embedding domain-specific 
knowledge into the messages being passed, we are able to obtain higher-quality solutions  
compared to general-purpose approaches, e.g., based upon Markov decision processes, 
which need to blindly ``learn'' the problem structure.

\section{System Model}\label{sec:system}

\begin{figure}
\centering
\includegraphics[width=0.45\textwidth]{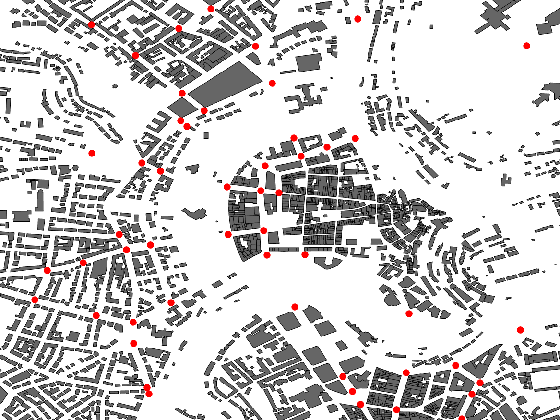}
\caption{\label{fig:scenario} Real-world scenario: Luxembourg city
  center. The red circles represent the locations of the traffic
  lights, i.e., of the gNBs.}
\end{figure}

To develop a system model that captures all the main aspects of a mmWave vehicular network, 
we consider a reference scenario based on real-world
mobility and infrastructure traces, as per~\cite{lust,tust}. Such traces
contain information about the topology of the city of Luxembourg, the road layout
(e.g., regulated intersections), as well as the mobility traces of
around several thousands of vehicles traveling within the city center,
accumulated over a 12-hour window. 
Based on this data, we construct a scenario as the one depicted in
Fig.~\ref{fig:scenario}, in which a set of gNBs, denoted by $\Gc$, are
co-located with traffic lights to serve a set of vehicles, i.e., the
mmwave {\em{users}}.    The service 
 is divided into a set~$\Zc$ of discrete
zones: each vehicle is, at any given time, into exactly one zone.
Below, we detail the characteristics of our scenario, and present the 
assumptions we make to build an accurate, yet tractable, system model.

\paragraph*{Antenna arrays and spatial signatures}
We assume that all network nodes
(gNBs and vehicles) have the same height and consider a 2D 
network topology, which allows for a simpler
mathematical analysis and a lighter notation. Indeed, while easily
generalizable to 3D, the 2D network model is already able to describe
and capture all important features of the system.
Both gNBs and vehicles are equipped with a uniform planar array (UPA)
of antennas, with the vehicle UPA being capable of analog beamforming
only. We assume the elements of a UPA arranged in a
square grid and spaced by $\lambda/2$, with $\lambda$ denoting the
signal wavelength. In particular, the gNB's UPA has size $N_t\times N_t$ elements, while the
vehicle's UPA has size $N_r\times N_r$ elements.
The surface of a generic UPA is vertically placed;
thus, the vector normal to the surface points to the horizon and has
azimuth $\psi$ (measured with respect to a global coordinate system).

If the UPA has $N\times N$ antenna elements, then its spatial signature in the
direction, defined by the azimuth $\phi$ (measured with respect to its
normal), is the $N^2$-size vector
\begin{equation} \sv(N,\phi)=  \onev_N \otimes \widetilde{\sv}(N,\phi)\label{eq:s}\,,
\end{equation}
where $\onev_N$ is a vector of length $N$ with all elements equal to 1,
$\otimes$ denotes the Kronecker product, and the $n$-th component of
$\widetilde{\sv}(N,\phi)$ is given by
$[\widetilde{\sv}(N,\phi)]_n = \ee^{\jj \pi n\sin \phi}$.

\paragraph*{Beamforming}
A beam, $b$, is generated by a gNB using a subset of antenna elements from
its UPA. In practice, the beam is obtained by coordinating (phasing)
the signals emitted by the antenna elements so that they globally act
as a single directional antenna whose main radiation lobe is
characterized by a specific half-power beam width (HPBW), $\alpha$,
and direction $\delta$. Beams with different HPBW can be obtained by
varying the number of antennas involved in the beam generation.
Specifically, if the array generating the beam $b$ has size $N\times N$
elements, the HPBW, $\alpha$, depends on $N$ through the relation:
\begin{equation}
  \alpha\approx \frac{1.78}{N}[\mbox{rad}] \approx \frac{102}{N}^\circ\,. \label{eq:hpbw}
\end{equation}
The direction of the beam can be set by properly choosing the {\em
  beamforming vector} $\vv$ i.e., the vector of phases applied to the
array elements in order to emit the considered beam, given by
\begin{equation} 
\vv \triangleq \frac{1}{N} \sv(N,\varphi) 
\label{eq:vv}
\end{equation}
where the angle $\varphi$ represents the direction of the generated
beam, in the azimuth plane, with respect to the normal to the gNB
UPA. In practical systems, the values that the angle $\varphi$ can take
are limited to a discrete set.  The direction of the beam in a global
horizontal coordinate system is then defined by the angle
$\delta = \varphi+\psi$, where we recall that $\psi$ is the azimuth of
the normal to the gNB UPA.

We denote by $\Bc$ the set of possible beams, which is common to all gNBs and 
whose cardinality is equal to the
number of possible beam directions multiplied by the number of HPBW
values available at the gNB. The set $\Bc$ contains also an extra element, i.e., the null-beam
denoted by $\emptyset$. In the following, the notation $b=\emptyset$
denotes that the beam $b$ is not emitted.

Let us define as $B$ the maximum number of beams that can be activated at a 
gNB $g$ ($g=1,\ldots, G$),  and denote with vector $\bv_g=[b_{g,1},\ldots,b_{g,B}]$ the 
generic beam-configuration adopted at gNB $g$.  
The $i$-th beam, $i=1,\ldots,B$, 
is characterized by direction $\delta_{g,i}$ and HPBW
$\alpha_{g,i}$. However, not all beam-configurations $\bv_g$ are possible
since two main constraints need to be accounted for.  First, 
the sum of the number of antennas elements simultaneously involved in the beam
generation should not exceed the number of antenna
elements of the gNB's UPA (i.e., $N_t^2$). Second, we must ensure that beams do not
overlap with each other, i.e., for every two beams $b_{g,i}, b_{g,j} \neq \emptyset$
simultaneously emitted on the same frequency band by gNB $g$, the following condition must
hold:
\begin{equation}
\left|\delta_{g,i}-\delta_{g,j}\right|\geq\frac{\alpha_{g,i}
+\alpha_{g,j}}{2}.
\end{equation} 
We then define
$\Fc \subseteq \Bc$ as the set of feasible beam configurations at a generic gNB.  
Finally, a network beam configuration, $\Bm$, can be described as an array of
$G$ beam-configurations, one for each gNB, i.e.,
$\Bm=[\bv_1,\ldots, \bv_g,\ldots\bv_G] \in \Fc^G$.

\paragraph*{Transmitted signal}
Radio resources available for communication are organized into an {\color{black} $N_b \times N_t$}
frequency-time matrix, whose time size is called frame.  A single
element of such matrix, named resource block, is characterized by a
bandwidth $W$ and a {\color{black} time fraction $\tau=1/N_t$. The total
bandwidth assigned for communication is then $N_bW$}.  
We assume that every beam transmits over all
resource blocks; therefore, given a network beam configuration, $\Bm$, the
baseband signal transmitted by the $i$-th beam of gNB $g \in \Gc$, in 
resource block $q$ can be modeled by vector:
\begin{equation}\label{eq:t}
  \tv_{g,i,q} = \vv_{g,i} x_{g,i,q}
\end{equation}
{\color{black} where $\vv_{g,i}$ is the beamforming vector in~\eqref{eq:vv}
particularized to beam $b_{i,g}$, and $x_{g,i,q}$ is a complex random
symbol with zero mean. Further, assuming a uniform power allocation over all resource blocks, 
we associate to $x_{g,i,q}$ a power equal to: $\EE[|x_{g,i,q}|^2] = P_{g,i}/N_b$.} Note
that the available transmit power at gNB $g$, $P_g$, is shared among
the beams simultaneously emitted therein. Therefore, the values
$P_{g,i}$ are subject to power allocation constraints that also
depend on the adopted beam configuration.

\paragraph*{Mmwave communication channel}
In a typical mmwave urban scenario, the channel between a gNB $g$ and
a zone $z$ can be modeled as described
in~\cite{rappaport-chanmodels, akdeniz-mmwave, 3gppchanmodel}.  Such
models consider $L_g(z)$ clusters of paths, each described by a complex
coefficient, $h_{g,\ell}(z)$, and two angles, $\phi_{g,\ell}(z)$ and
$\theta_{g,\ell}(z)$, $\ell=1,\ldots, L_g(z)$, which represent, respectively, the departure
and arrival direction of the signal, measured with respect to the normal to the
transmitting and receiving UPAs.  Since we consider zones to be sufficiently small
so that vehicles therein (if any) experience the same propagation
channel for a given beam and gNB, in the following we associate a zone
with a unique UPA and receiver, and we detail the channel model by
referring to a gNB-zone communication link.
For a given network beam configuration, $\Bm$,
the channel experienced by beam $b_{g,i}$, connecting gNB $g$ to zone $z$, is given by the matrix:
\begin{equation}\Hm_{g,i}(z) = \sqrt{\frac{1}{L_g(z)}}\sum_{\ell=1}^{L_g(z)} h_{g,\ell}(z)
    \uv_{g,\ell}(z){\muv_{g,i,\ell}(z)}\Herm\label{eq:H} 
\end{equation}
where $\muv_{g,i,\ell}(z)\triangleq \sv(N_{g,i}, \phi_{g,\ell}(z))$
and $\uv_{g,\ell}(z)\triangleq \sv(N_r, \theta_{g,\ell}(z))$ are the
signatures of the transmit and receive antenna arrays, and $N_{g,i}^2$
is the number of antenna elements used by beam $b_{g,i}$.  The channel
model in~\eqref{eq:H} does not depend on the resource block $q$, hence
it is assumed to be frequency flat and characterized by a coherence
time larger than the frame time.

\paragraph*{Received signal}
For a given network beam configuration $\Bm \in \Fc^G$, the signal
carried by beam $b_{g,i}$ in resource block $q$ and received within
zone $z$ can be represented by the following $N_r^2\times 1$ vector:
\begin{equation}\label{eq:Y}
  \yv_{g,i,q}(z) = \Hm_{g,i}(z) \tv_{g,i,q}+ \etav_{g,i}(z)
\end{equation}
where $\Hm_{g,i}(z)$ and $\tv_{g,i,q}$ are given by~\eqref{eq:H}
and~\eqref{eq:t}, respectively, {\color{black} and $\etav_{g,i}(z)$ is a term
accounting for noise and interference in zone $z$, e.g., caused by
beams generated by nearby gNBs (in general, all beams in $\Bm$
except for $b_{g,i}$).  We assume that vector $\etav_{g,i}(z)$ has
complex Gaussian independent entries with zero mean and covariance
$\EE[\etav_{g,i}(z) {\etav_{g,i}(z)}\Herm]=(N_0W+I_{g,i}(z))\Id$}
where $N_0$ is the thermal noise power spectral density, $W$ is the
signal bandwidth, and $I_{g,i}(z)$ is the interference power.  Since
only analog beamforming is possible at the receiver, the receiver
applies to $\yv_{g,i,q}(z)$ the vector of weights $\wv(z)$ given by
$ \wv(z) \triangleq \frac{1}{N_r}\sv(N_r, \varphi^{(z)})$.  Notice
that such vector has norm 1 and allows the receiver to generate a beam
in the direction specified by the angle $\varphi^{(z)}$.  After
weighting the UPA output, the receiver obtains {\color{black}
  \begin{equation}
    \wv\Herm (z) \yv_{g,i,q}(z) = \widetilde{h}_{g,i}(z) x_{g,i,q} + \wv\Herm(z)\etav_{g,i}(z) \label{eq:tildeh} 
  \end{equation} 
  where the scalar}
  $\widetilde{h}_{g,i}(z) = \wv\Herm(z) \Hm_{g,i}(z) \vv_{g,i}$
  represents the overall communication channel summarizing the effects
  of the signal propagation and of the antenna and beam design.  The
  term {\color{black} $\wv\Herm (z)\etav_{g,i}(z)$} is a Gaussian random variable
  with zero mean and variance
  $N_0W+I_{g,i}(z)$, and the
  received signal power is given by {\color{black}
  $\EE[|\widetilde{h}_{g,i}(z)x_{g,i,q}(z)|^2]=P_{g,i}|\widetilde{h}_{g,i}(z)|^2/N_b$}. Note
  that the values of the transmit power, $P_{g,i}$, and of channel 
  $\widetilde{h}_{g,i}(z)$ depend on the specific
  network beam configuration, $\Bm$. To stress this dependence, in the
  following,  we will write
  $P_{g,i}(\Bm)$ and $\widetilde{h}_{g,i}(z,\Bm)$.
  
 \paragraph*{Serving and interfering beams}

 Each zone $z\in \Zc$, is allocated a set of resource blocks,
 $\Qc(z)$, and a set of serving beams. In turn, a beam can serve one or
 more zones, depending on its direction and HPBW.
 Specifically, for a given network beam configuration, $\Bm$, the set
 of beams {\em serving} zone $z$ is
 denoted by $\Sc(z,\Bm)$.  In other words, the elements of
 $\Sc(z,\Bm)$ are pairs $(g,i)$; and $(g,i)\in \Sc(z,\Bm)$ if, under
 network beam configuration $\Bm$, the $i$-th beam emitted by gNB $g$
 serves zone $z$.  Note that, in CoMP-like communications, we can have
 $1\le |\Sc(z,\Bm)|\le G_c$ where $G_c$ is the maximum number of gNBs
 that can partake in the coordinated transmission. Clearly, when no
 CoMP is enabled, $G_c=1$.

 Similarly, $\Ic(z,\Bm)$ denotes the set of 2-tuples $(g,i)$
 identifying the beams $b_{g,i}$ {\em interfering} in zone $z$, under
 network beam configuration $\Bm$.  By using these definitions, the
 received signal and interference powers {\color{black} for any resource block $q\in \Qc(z)$ in zone $z$
 can be written, respectively, as
 \begin{eqnarray}
   \widetilde{P}(z,\Bm) &=&\left| \sum_{(g,i)\in \Sc(z,\Bm)} \sqrt{\frac{P_{g,i}(z,\Bm)}{N_b}} \widetilde{h}_{g,i}(z,\Bm)\right|^2 \non
               I(z,\Bm) &=& \sum_{(g,i)\in \Ic(z,\Bm)} \frac{P_{g,i}(z,\Bm)}{N_b}|\widetilde{h}_{g,i}(z,\Bm)|^2 \nonumber
 \end{eqnarray}
and the achievable rate as
   \begin{equation}\label{eq:R_qzB}
    R(z,\Bm) = W\tau\log_2\left(1+\frac{\widetilde{P}(z,\Bm)}{N_0W+I(z,\Bm)}\right)\,.   
  \end{equation}
Given $\Bm$, the total network data rate is given by:
   \begin{equation}\label{eq:R_B}
    T(\Bm) = \sum_{z\in \Zc} |\Qc(z)| R(z,\Bm)\,. 
  \end{equation}
}

\begin{table}[t]
\caption{{\color{black}Main notation\label{tab:notation}}}
\begin{tabularx}{1\columnwidth}{|l|X|}
  \hline
  {\bf Variable} & {\bf Description} \\ \hline \hline
  $\Gc$ & set of gNBs \\ \hline
  $\Zc$ & set of zones \\ \hline
  $\Qc(z)$ & set of resource blocks assigned to zone $z$ \\ \hline
  $\Fc\subseteq \Bc$ & set of feasible beam configurations in a gNB\\ \hline
  $\Bm$ & network beam configuration \\ \hline
  $\bv_g$ &  beam configuration at gNB $g$\\ \hline
  $\alpha_{g,i}$ & HPBW of the $i$-th beam when the local configuration~$\bv_g$ is adopted at gNB~$g$   \\ \hline
  $\delta_{g,i}$ & direction of the $i$-th beam when the local configuration~$\bv_g$ is used at gNB~$g$  \\ \hline
  $\Pi(\Bm)$ & probability that the network-wide configuration~$\Bm$ is adopted\\ \hline
  $\pi_g(\bv_g)$ & probability that the local configuration~$\bv_g$ is adopted at gNB~$g$\\ \hline
  $T(\Bm)$ & network data rate achieved under beam configuration $\Bm$\\ \hline
  $\chi^{(g,h)}(\bv_g,\bv_h)$ & joint compatibility function between gNBs $g$ and $h$ when configurations $\bv_g$ and 
  $\bv_h$ are adopted\\
  \hline
\end{tabularx}
\end{table}

\section{A Randomized Approach to Network Throughput Maximization}\label{sec:problem-formulation}

Given the set of gNBs, $\Gc$, the maximum number of
  supported beams at each gNB, $B$, and the set of zones, $\Zc$, our
  goal is to determine the best beam configuration to be used at each gNB.  More
  specifically, we aim at jointly addressing the following questions
  while maximizing the overall network data rate:
\begin{itemize}
\item[{\em (i)}]how many beams, of what width and direction, each gNB should set up, and  
\item[{\em (ii)}] which zones should be associated to which gNB, and scheduled on which beam. 
\end{itemize}

Owing to the highly dynamic scenarios we target and to the stochastic nature of the wireless channel, 
we adopt a {\em randomized approach}, whereby (i) decisions concern the {\em probability} with which a given 
configuration is adopted, and (ii) the actual configuration to enact is chosen according to such probabilities. 
Randomized approaches similar to ours have been long used in many fields, including 
routing~\cite{flury2008randomized,zhao2020can}, resource allocation~\cite{rost2019virtual}, 
and security~\cite{conti2007randomized,shu2010secure}.
It follows that, in our problem formulation (i) the decision variables correspond to 
{\em probabilities} that each beam configuration is adopted, and (ii) 
the objective function is the {\em expected value} of the actual target metric. 
Below, first we present a centralized formulation and discuss its complexity, then 
we introduce a distributed version of the problem that allows for local decision 
making based on local information. The main notation we use in our formulations is summarized in
\Tab{notation}.

\paragraph*{A centralized formulation}

Let us first formulate a centralized  optimization problem  that aims at a globally optimal solution. Such a
problem needs to be time-dependent and solved
periodically\footnote{The extension to an event-driven decision making
  process is however straightforward.}. Upon  
solving the problem, the solution is represented by the probability values associated to each beam configuration, 
which are delivered to the set of gNBs. The gNB nodes set their beams according to the received configuration probabilities, 
by devoting to each of them a fraction of time that is proportional to the probability values.  
Since the problem
formulation holds at every decision period, to simplify the notation,
in the following we do not highlight the dependency on time.

Let $\Pi(\Bm)$ represent the probability that network 
beam configuration $\Bm$ is selected. Then the total network data rate is given by:
\begin{equation}\label{eq:T_average}
\EE_\Pi[T] =  \sum_{\Bm \in \Fc^G}\Pi(\Bm)T(\Bm)
\end{equation}
and our problem can be formulated as:
\begin{equation}\label{eq:max_T_average}
\max_{\Pi(\Bm)} \EE_\Pi[T(\Bm)] \,.
\end{equation}
Note that~\eqref{eq:max_T_average} is
maximized when $\Pi(\Bm) = 1$ for $\Bm=\Bm^{\rm max}$ and 0 otherwise,
where $\Bm^{\rm max}= \arg\max_{\Bm}T(\Bm)$. The optimal beam
configuration $\Bm^{\rm max}$ is however difficult to compute due to
the interference among beams, the need for centralized global knowledge, and the non-linear nature of the
problem. Indeed, non-linear problems are
notoriously complex to solve, and solution strategies only find
local optima in the general case.

\paragraph*{A decentralized formulation}
In light of the issues above, we are interested in a decentralized solution 
where each gNB independently chooses
its own beam configuration according to a local probability distribution $\pi_g(\bv_g)$.  
To this end, we consider as objective function the following expression:
\begin{equation}\label{eq:T_average_decentralized}
\EE_{\pi_1, \ldots, \pi_G}[T(\Bm)]\mathord{=}  \sum_{\bv_1\in \Fc}\cdots \sum_{\bv_G\in \Fc} 
\prod_{g=1}^G\pi_g(\bv_g)T([\bv_1,\ldots,\bv_G])
\end{equation}
and write our problem as:
\begin{equation}\label{eq:max_T_average_decentralized}
\max_{\{\pi_g\}} \EE_{\pi_1, \ldots, \pi_G}[T(\Bm)] \,.
\end{equation} 
Under this decentralized approach, once a decision is made, each gNB $g$ will 
implement the optimal policy by applying in each of the $K$ time steps one of 
the possible configurations, with such probability as specified by 
the densities $\pi_g(\bv_g)$. 
Importantly, the following holds:
\begin{lemma}
The joint probability obtained using the solutions to 
(\ref{eq:max_T_average_decentralized}) coincides  with the optimal solution to  
(\ref{eq:max_T_average}).
\end{lemma}
\begin{IEEEproof} 
The proof comes from the well-known fact that, if variables are independent, 
then the joint distribution of a set of random variables is equal to the product 
of individual, marginal distributions. In our scenario, independence is guaranteed by the 
fact that each individual gNB~$g$ performs its own randomized strategy, i.e., 
chooses the concrete strategy to enact according to~$\pi_g$, with no influence from other gNBs. 
Then, it is sufficient to observe that 
the marginal densities maximizing~\eqref{eq:T_average_decentralized}
are given by $\pi_g(\bv_g) = 1$ for $\bv_g=\bv_g^{\rm max}$ and 0 otherwise, for $g=1,\ldots,G$
and $[\bv_1^{\rm max},\ldots,\bv_G^{\rm max}]=\Bm^{\rm max}$, as defined in the centralized approach. 
\end{IEEEproof}

Moving from the maximization over the joint distribution in~\eqref{eq:max_T_average} to the one over the marginal 
distributions in~\eqref{eq:max_T_average_decentralized} does not, {\em per se}, change the complexity of 
the problem, nor its solution. However, it does provide us with valuable insights on a possible solution 
strategy, namely, one leveraging {\em belief propagation} (BP) algorithms.

In general, BP allows multiple agents to cooperatively estimate the marginal distributions of a set of 
random variables. Specifically, each agent is associated with a random variable, and an {\em influence graph} 
expresses which variables (hence, which agents) influence one another. The algorithm works 
iteratively, with agents that are neighbors on the influence graphs exchanging messages 
(indeed, BP belongs to the family of {\em message passing} algorithms). 
BP is guaranteed to converge to the centralized solution if the influence graph is a tree, 
but works remarkably well under a much wider set of conditions.

In our case, agents correspond to gNBs, the random variables to estimate are the local 
decisions~$\pi_g(\bv_g)$, and messages express the extent to which decisions of different gNBs 
conflict with one another. By leveraging BP, we are able to make swift, high-quality decisions that 
(i) for particular network topologies, {\em match} the optimal ones, i.e., the ones we would obtain 
by solving~\eqref{eq:max_T_average}, and (ii) in general, are very close to the optimal decisions 
in most practical cases.

\section{CRAB: Coverage-Rate Aware Belief Propagation\label{sec:solution}}

Our solution strategy, named  Coverage-Rate Aware Belief propagation (CRAB), is predicated 
on allowing each gNB to make {\em
  local} decisions about the number, direction, and width of its
beams. Such decisions are aimed to avoid {\em conflicts} among distributed beam
management decisions, thus yielding high serving rate and coverage to vehicular users. 
Indeed, in our scenario, improving the coverage of users is a very good way towards optimizing 
the total data rate~\eqref{eq:R_B}, as good coverage implies less interference and more served users.

As discussed above, we follow a {\em randomized} approach, where the
decision variables are represented by probability distributions;
specifically, we associate to each gNB~$g$ and beam
configuration~$\bv_g$ a probability~$\pi_g(\bv_g)$. Then, at
every time step $k$, each gNB adopts one of the possible
configurations with probability $\pi_g(\bv_g)$.  Such decisions
are local in nature, therefore, we want to allow each gNB to choose
the distributions $\pi_g(\bv_g)$ that maximize its performance.  
At the same time, we
have to avoid {\em conflicts} among decisions made by different gNBs.
We define a conflict as two or more beams of distinct gNBs covering
the same zone(s): such a situation is doubly wasteful, as (i) beams
may interfere with each other, and (ii) one of the them could cover
different, hitherto unserved, zones instead.  Note that the system
would greatly benefit from anticipating such conflicts and avoiding
them, as detecting a conflict a posteriori would entail that the
involved gNBs have to change their configuration so as to remove the
conflict itself, with the intrinsic overhead as well as the risk to
create a new conflict.

\begin{figure*}
\centering
\includegraphics[width=0.75\textwidth]{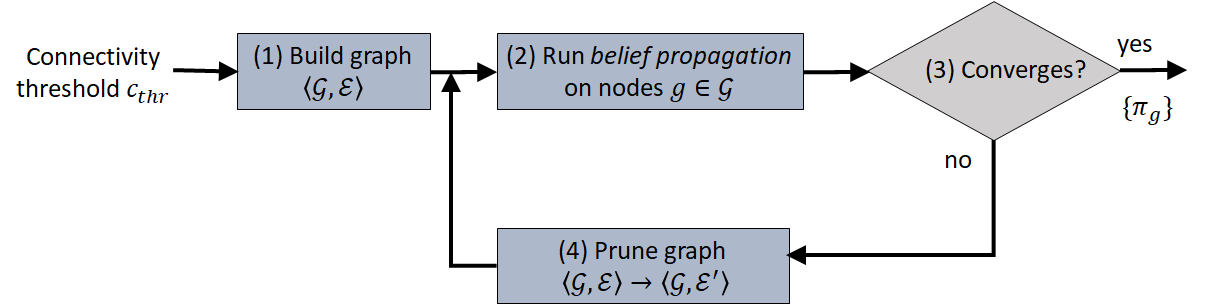}
\caption{\label{fig:flowchart} Schematic representation of the CRAB framework: 
(1) the interaction graph is built using the given connectivity threshold as detailed 
in Sec.~\ref{subsec:graph}; (2) the {\em{belief propagation}} algorithm is run on the nodes of the graph, 
as described in Sec.~\ref{subsec:algo}; (3) if the algorithm does not converge within a maximum number of 
iterations, in step (4)  the graph is pruned as described in Sec.~\ref{subsec:prune}. 
Steps (2)-(4) are repeated until convergence is reached and the marginal probabilities $\pi_g$ are obtained. }
\end{figure*}

We tackle this conundrum by:
\begin{itemize}
\item creating an {\em interaction graph}, modeling the mmwave network and 
summarizing which gNBs'
  decisions may conflict with each other;
\item apply a belief propagation algorithm to solve the distributed problem
  introduced in Sec.\,\ref{sec:problem-formulation};
\item if the algorithm does not converge, {\em pruning} such a graph
  and re-applying the belief propagation algorithm until convergence
  is reached and a solution is obtained.
\end{itemize}
{\color{black}
These steps are depicted in the general scheme of the CRAB process in \Fig{flowchart}.
}

Below, we first present how the interaction graph is built, in such a
way that the mmwave network characteristics are accounted for  (Sec.\,\ref{subsec:graph}).  Then, we
associate with each gNB a {\em state}, i.e., a beam configuration, and,
given the objective in~\eqref{eq:max_T_average_decentralized}, we design a belief propagation
algorithm which applies the message passing approach and yields a
probability distribution over the local beam configuration decisions
(Sec.\,\ref{subsec:algo}).  Finally, we detail how the interaction
graph can be pruned if necessary, so as to make the algorithm reach
stable and mutually beneficial solutions to be adopted at the gNBs
(Sec.\,\ref{subsec:prune}).

\subsection{Building the interaction graph}\label{subsec:graph}
\begin{figure*}
\centering
\includegraphics[width=0.38\textwidth]{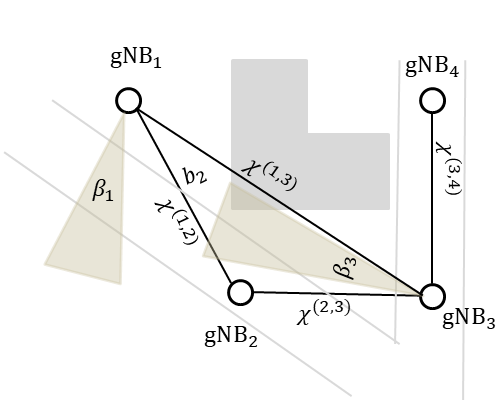}
\hspace{2mm}
\includegraphics[width=0.38\textwidth]{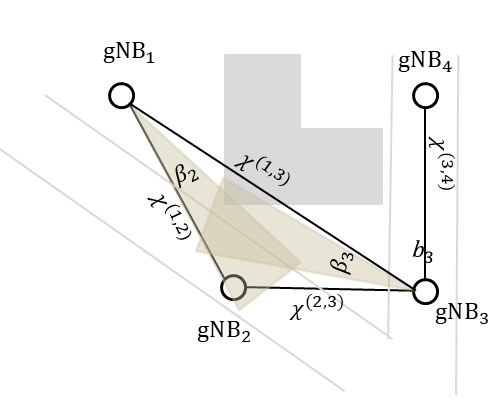}
\caption{\label{fig:graph_model}
Examples of interaction graph modeling, 
highlighting two possible beam configurations at each gNB (i..e, graph node).
}
\end{figure*}

To address the beam management problem, we model the network as a
graph, composed of a set of nodes coinciding with $\Gc$, i.e., with
each node representing a single gNB, and a set of edges $\Ec$ between
said nodes. As exemplified in Fig.~\ref{fig:graph_model}, an edge
$e_{g,h}$ between nodes $g$ and $h$ exists if the value of the
connectivity criterion between the two vertices, $c_{g,h}$, is above
a certain threshold $c_{\rm thr}$, i.e.,
\[\Ec=\{e_{g,h}: \forall g, h\in\Gc \wedge c_{g,h}>c_{\rm thr}\}\,.\]
As detailed below, such criterion should reflect the gNB's potential
to interfere with each other's transmissions and can include, but it is
not limited to, the inter-gNBs distance, the existence of line-of-sight (LoS) conditions
between gNBs, or the gNBs coverage overlapping in terms of number of users.

\begin{figure*}
\centering
\includegraphics[width=0.31\textwidth]{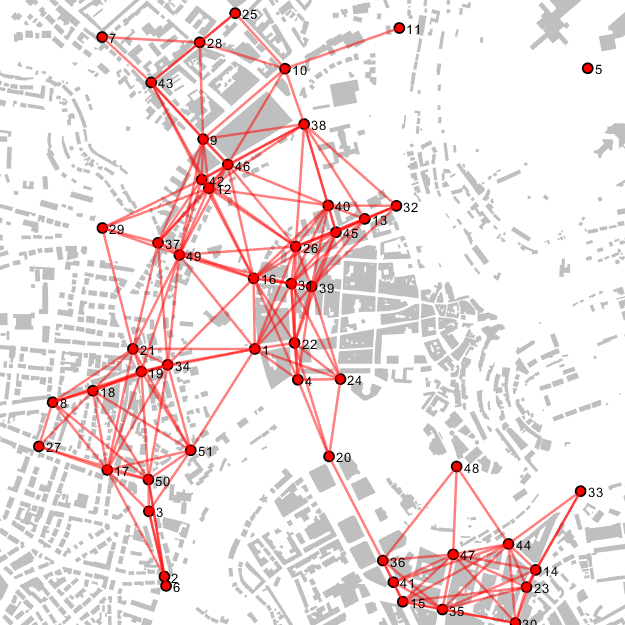}
\hspace{2mm}
\includegraphics[width=0.31\textwidth]{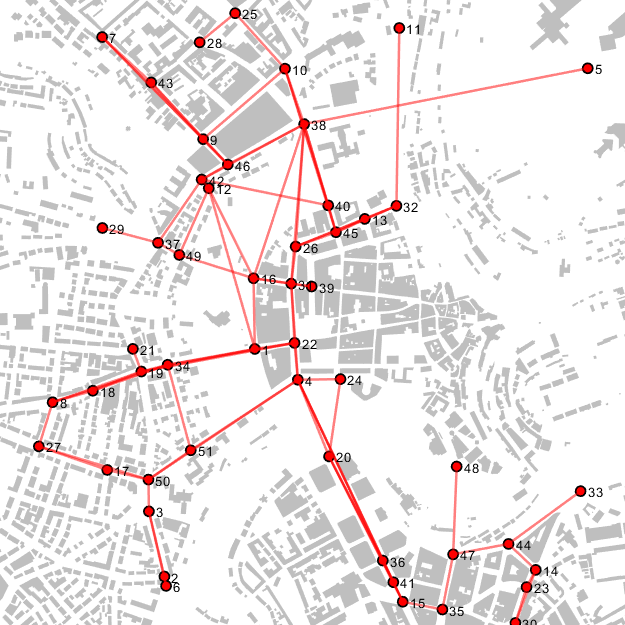}
\hspace{2mm}
\includegraphics[width=0.31\textwidth]{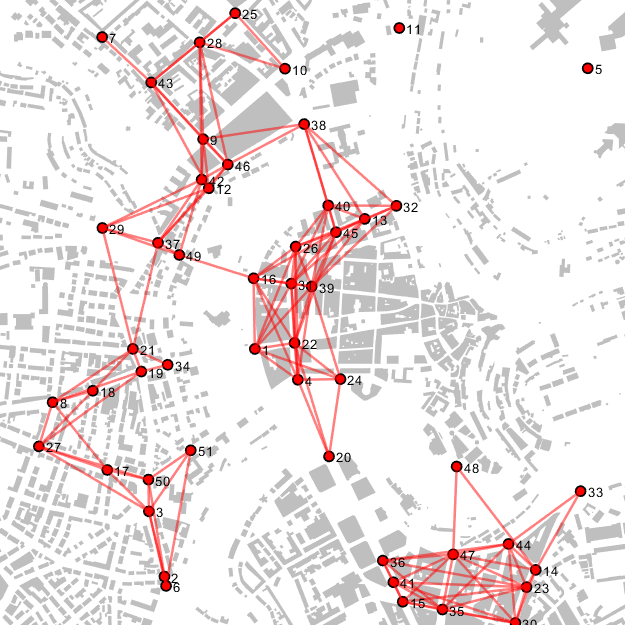}
\caption{\label{fig:graphvarconn}Mmwave network graph using different connectivity criteria: 
(left) distance-based criterion with $c_{g,h}=\frac{1}{d_{g,h}}$ and $d_{g,h}=400$\,m; (center) LoS-based criterion; 
(right) overlapping coverage criterion with $c_{\rm thr}=0.1$.}
\end{figure*}

Examples of how the graph of the mmWave network in Luxembourg city would look like  
using different criteria  
is shown in \Fig{graphvarconn}. In the leftmost figure, 
the value of the connectivity criteria is expressed as a 
function of the distance between the nodes, i.e., $c_{g,h}=\frac{1}{d_{g,h}}$, and the threshold is 
related to a fixed distance $c_{\rm thr}=\frac{1}{d_{\rm thr}}$ where 
$d_{\rm thr}=400$~m, as it has been shown  that 
the probability of maintaining a mmwave communication link  over longer distances is negligible 
\cite{akdeniz-mmwave}.  
In the center figure, the criteria is the existence of LoS between two nodes, which can be 
expressed as $c_{g,h}={\rm LoS}(g,h)$, where ${\rm LoS}(g,h)$ is a binary variable indicating whether 
there is LoS between the two nodes, and $c_{\rm thr}=0$. Finally, in the right-hand plot 
the connectivity criteria is based on the fraction of the number of vehicles under 
the coverage of  the two gNBs with respect to the total number of users covered by the two gNBs separately, and the threshold values has been 
 set to $c_{\rm thr}=0.1$.

The various
criteria can also be combined, and identifying the most suitable
criterion to be considered is one of the goals of this work.  It
should be noted, however, that while the first two criteria are topology
based, and we can assume that the structure of the graph will be
fixed, the third one depends on vehicle traffic patterns, therefore
the structure of the graph might change over time, as traffic patterns
shift.

\subsection{Configuring the belief propagation algorithm}\label{subsec:algo}
Belief propagation is an algorithm allowing to
infer the (local) marginal distributions of a set of random variables
taking into account their mutual correlation. 
The algorithm is defined over graphs where each random variable is associated to a node, and
works by letting {\em messages} flow along the graph edges. A message is
a real valued function that measures the influence that a random variable (i.e., a node)
exerts on the neighboring ones. The algorithm works iteratively and, at every iteration, 
the marginal distributions of the random variables at each node are computed and updated,
until convergence is reached.

In our scenario the graph is the network interaction graph obtained under one of
the connectivity criteria mentioned in Sec.~\ref{subsec:graph}.  The random variables
are the beam configurations at the nodes, $\bv_g$ ($g=1,\ldots,G$), and their marginal distributions
are the $\pi_g(\bv_g)$'s previously introduced.

\paragraph*{Joint compatibility function}
The interaction among random variables is described by the joint
compatibility function $\chi^{(g,h)}(\bv_g,\bv_h)$, which, for every
pair of nodes $g,h\in \Gc$, measures the compatibility between beam
configurations $\bv_g$ and $\bv_h$ when they are simultaneously
activated by nodes $g$ and $h$, respectively.

We design our compatibility function based upon the intuition that two configurations {\em interfere} 
with each other when they cover the same set of users. Such a situation hurts the objective 
in~\eqref{eq:max_T_average_decentralized} in two ways, namely:
\begin{itemize}
    \item it results in fewer users being served, hence, the total data rate decreases;
    \item it creates interference for the users that do get served, further reducing the total data rate.
\end{itemize} 
For the above reasons, improving {\em coverage} is strongly linked with increasing the total data rate 
in (\ref{eq:R_B}).

For example, if we look at
Fig.~\ref{fig:graph_model} where {\color{black} the maximum number of activated beams is $B=1$, we expect that the
compatibility value between the beam configurations $b_1=\beta_1$ at node
$gNB_1$ and beam $b_3=\beta_3$ at node $gNB_3$, to be higher than the
value obtained when $b_1=\beta_2$ at node $gNB_1$ and $b_3=\beta_3$ at
node $gNB_3$, i.e., $\chi^{(1,3)}(\beta_1,\beta_3)>\chi^{(1,3)}(\beta_2,\beta_3)$}, since in the first case the beams interfere with each
other. 
Accordingly, we define the  joint compatibility function as:
\begin{equation}
\label{eq:compat}
  \chi^{(g,h)}(\bv_g,\bv_h) = \sum_{z\in\Zc}\sum_{q\in\Qc(z)}R_q(z,\Bm^{(g,h)})
\end{equation}
where\footnote{Without loss of generality, we assume $g<h$.}
$\Bm^{(g,h)} =
[\boldsymbol{\emptyset},\ldots,\boldsymbol{\emptyset},\bv_g,\boldsymbol{\emptyset},\ldots,\boldsymbol{\emptyset},\bv_h,\boldsymbol{\emptyset},\ldots,\boldsymbol{\emptyset}]$,
$\boldsymbol{\emptyset}=[\emptyset, \ldots, \emptyset]\in \Fc$ is the
null beam set, and $R_q(z,\Bm)$ is the rate defined
in~\eqref{eq:R_qzB}.  In practice, $\chi^{(g,h)}(\bv_g,\bv_h)$ returns
the rate achieved by the network when beam-configurations $\bv_g$ and
$\bv_h$ have been selected at gNBs $g$ and $h$, respectively, and all
other gNBs are silent.
It is interesting to observe how the compatibility function in~\eqref{eq:compat} is not directly 
derived from the objective~\eqref{eq:max_T_average_decentralized}; rather, it incorporates domain- 
and scenario-specific knowledge about which situations ought to be avoided in order to improve performance.

\begin{figure}
\centering
\includegraphics[width=0.4\textwidth]{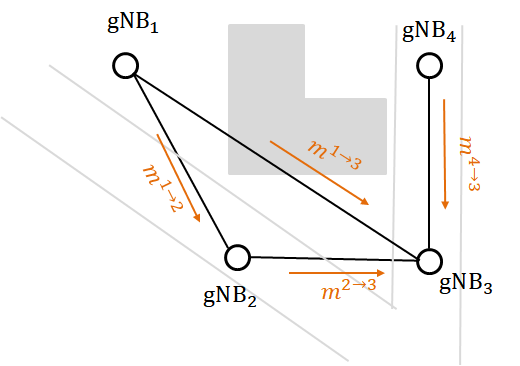}
\caption{\label{fig:messagepassing}An example of message passing in the CRAB framework.}
\end{figure}

{\bf Message passing process.}  As mentioned above, the belief
propagation algorithm works by exchanging messages along graph edges as shown in \Fig{messagepassing}.
Let us denote by $m^{g\to h}(\bv_h)$ the message that node $g$ sends to
node $h$ about  beam configuration, $\bv_h$. %

Once the interaction graph is built, we define $\nu(g)$ as the set of neighbors of node $g$ in the graph. 
We then design a belief propagation algorithm that, thanks
to the compatibility function defined above, yields probabilities $\pi_g(\bv_g)$'s that maximize 
the network data rate. 
Specifically, at each iteration, a node $g$ computes the messages to be sent to each of its neighbors, $h$, 
and for each of the neighbor's beam configuration $\bv_h$, according to:
\begin{equation}
m^{g\to h}(\bv_h)\mathord{=}\sum_{\bv_g\in\Fc} 
\chi^{(g,h)}(\bv_g, \bv_h)\hspace{-2mm}\prod_{k\in \nu(g)\backslash h}m^{k \to g}(\bv_g) \,\,\, \forall\,h,\bv_h
\end{equation} 
where  $m^{k \to g}(\bv_g)$ is the last message received by $g$ from $k$ about its beam configuration $\bv_g$. 
At the first iteration of the algorithm messages are initialized to a constant, arbitrary chosen value. 

In the first iteration all the nodes send their initialized messages to the respective outgoing nodes. In the following iterations, a node will send an outgoing message, once it has 
received all incoming messages necessary to compute 
it{\color{black}\footnote{We consider that inter-gNB communication is enabled through reliable links, 
hence no messages are lost during the exchange.}. An iteration is complete once the node calculates and 
sends all of its outgoing messages.}
Further, notice that since the joint compatibility function can take any value, we must ensure 
a proper normalization of the values $\pi_g(\bv_g)$ at each iteration of the algorithm. 
Such normalization is obtained by imposing: 
\[C_g = \sum_{\bv_g\in \Fc}\prod_{h\in\nu(g)}m^{(h\to g)}(\bv_g), \,,,\forall g=1,\ldots, G\,.\]

Convergence is reached when the difference
between all consecutively outgoing messages is negligible. If convergence is not reached within a 
maximum number of iterations, 
we proceed as explained in Sec.~\ref{subsec:prune}. 
Upon convergence, the marginal distributions of the beam configuration at node $g$ is given by:
\begin{equation}
\pi_g(\bv_g) = \frac{1}{C_g}\prod_{h\in\nu(g)}m^{(h\to g)}(\bv_g) 
\end{equation}
for all $\bv_g\in \Fc$.  
Each node can then locally decide regarding the beam configuration by
randomly selecting the state according to the obtained marginal
probability.

\subsection{gNB graph pruning}\label{subsec:prune}

Belief propagation does not guarantee
convergence in graphs that contain cycles \cite{ihlerloopy}, and, in general, the conflict graph
of a mmwave network is not guaranteed not to contain cycles.
The
application of belief propagation in such graphs, commonly referred to
as {\em{loopy belief propagation}}, is known to converge in most
practical cases, although no guarantee of convergence can be provided
\cite{ihlerloopy}. Rather, the existing relevant literature provides upper bounds
conditions under which the graph will converge.

It should be
noted, however, that loopy belief propagation can converge even if such conditions are
not satisfied. {\color{black}  In general, the convergence 
in such graphs depends heavily on the irregularity of the geometry of the graph itself and the variability 
in the strength, i.e., the 
range of the values of the compatibility function along the different edges, as shown in \cite{ihlerloopy}}.  
It follows that the criteria used to determine the connectivity between
the nodes plays a significant role. 
As already discussed, in the case of mmwave networks, several factors
can be taken into account when determining whether two gNBs are
connected; nonetheless, it is quite apparent that in an urban
scenario, loops are unavoidable.
 To ensure convergence, we therefore take the following
approach: if the graph does not converge within a maximum number of iterations, we prune the graph by using
Kruskal's reverse delete algorithm \cite{tar}, which removes the least
significant edge that does not affect the overall connectivity of the
graph. In the worst case, the pruning process leads to a tree-like
graph for which convergence is guaranteed, but our results (see Sec.\,\ref{sec:results}) show that
usually the graph converges much earlier than that.

\section{Numerical Results}\label{sec:results}

We evaluate the effectiveness of our approach by considering the real-world network 
layout of the Luxembourg city center, as described in \Sec{system}, 
and the realistic vehicular mobility trace in~\cite{lust}. 
The system parameters are configured as follows. The
center frequency available for mmwave communication is set to 
$52$~GHz, while the available bandwidth is $W=400$~MHz. The latter 
corresponds to the maximum allowed bandwidth in 5G New Radio (NR) 
using numerology $\mu=3$ with subcarrier spacing of $120$~kHz and 264 
resource blocks \cite{3gpp-bstrx}.  Further,  we assume that all gNBs are equipped with a $64\times64$
uniform planar array (UPA) with up to 4 RF chains transmitting at a maximum power 
of $P_g=33$~dBm, while
users are equipped with a $8\times8$ UPA with a single RF chain. 
To simulate the mmwave channel, we use the statistical approximation of the 3GPP channel model  accounting for the Doppler effect,
shadowing and multipath fading, and we set the large-scale
parameters used for modeling as in \cite{noi-isj20}.

Beam directions can take any integer value between $0^{\circ}$ and $360^{\circ}$, but 
we limit the number of possible beamwidth configurations to 
$\{5^{\circ}, 10^{\circ}, 15^{\circ}\}$. 
The CRAB algorithm is executed every second, and the total duration of the simulation is 10~s. 
We assume that all vehicles in the network are requesting data  for the entire duration of the simulation (the so-called full buffer model), and we focus on the downlink performance of the network. 
The resource allocation is performed according to the proportional 
fair algorithm, while user association is based on the 
strongest received reference signal. The effective data
rate for each vehicle is derived from the calculated SINR, 
by using the 4-bit channel quality indicator (CQI) table
in \cite{3gpp-tech}, which maps the reported CQI onto a particular modulation
coding scheme (MCS) and spectral efficiency value.
For the purposes of this study, the SINR to CQI mapping
has been performed using the spectral-efficiency based approach presented 
in~\cite{mezzavilla-cqi}.

\subsection{Connectivity analysis}

Due to the influence of the interaction graph on the overall performance of the CRAB 
algorithm, we begin by assessing how the topology of such a graph is influenced 
by the connectivity criterion used to determine whether or not to draw an edge 
between two gNBs. We compare two of the criteria discussed 
in Sec.\,\ref{subsec:graph}, namely, 
the distance between gNBs (hereinafter referred to as distance-based criterion) and  
the fraction
of the vehicles  that are covered by~{\em both}~$g$ and~$h$
(hereinafter referred to as coverage-based criterion).

It is important to point out that, regardless of the criterion employed, 
there is an inherent trade-off between the complexity of the CRAB 
algorithm (which in turn depends upon the connectivity of the interaction graph) 
and the performance of the resulting solution, i.e., the value of the objective \Eq{R_B}. 
We can therefore compare the two connectivity criteria we consider by 
characterizing the trade-offs that can be reached by adopting either of them.

\begin{figure}
\centering
\includegraphics[width=0.45\textwidth]{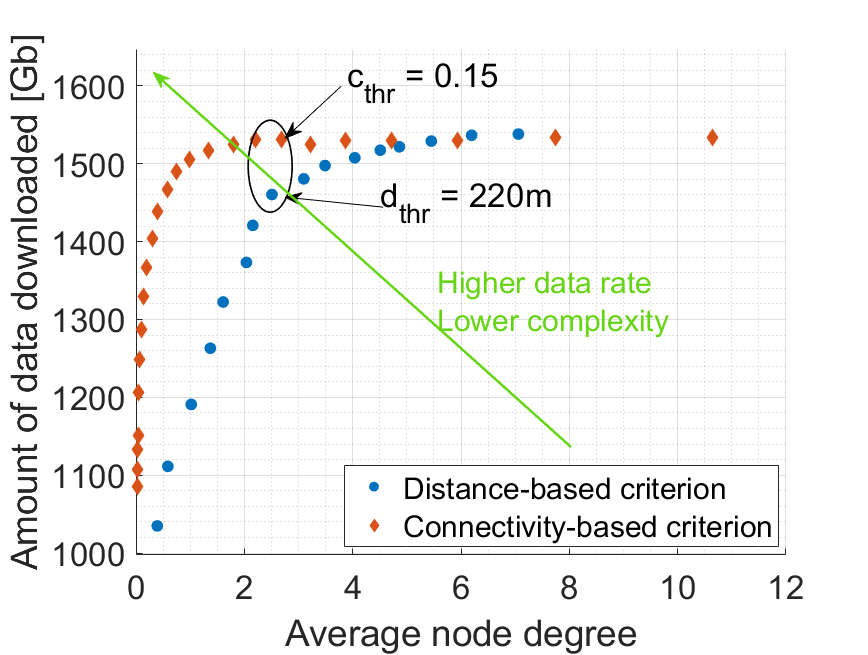}
\includegraphics[width=0.45\textwidth]{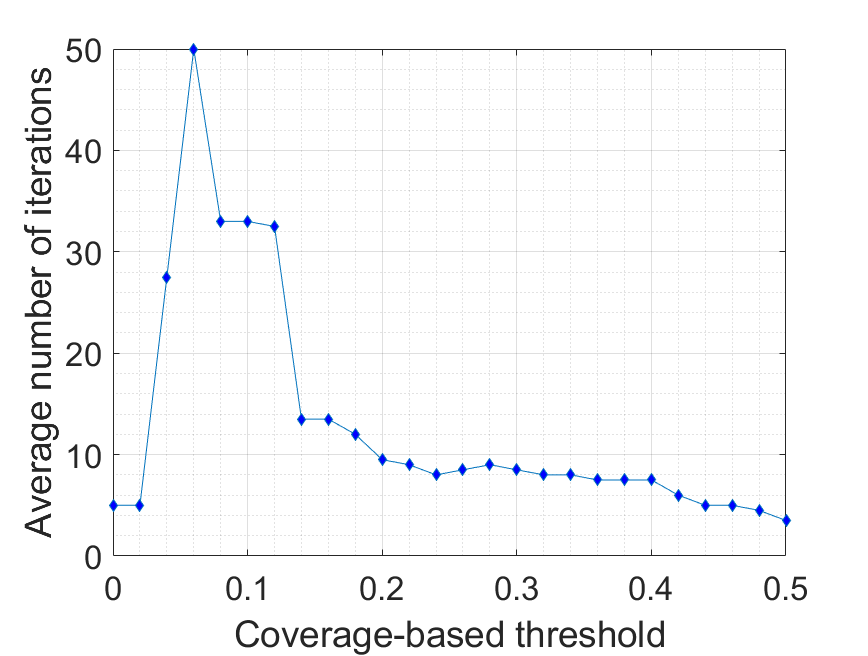}
\caption{\label{fig:graphvarconn_perf}Performance of the CRAB algorithm when using 
the distance-based criterion (blue) and the coverage-based criterion (orange): 
amount of data downloaded vs. average node degree (left); Average number of iterations as a function of the coverage-based connectivity threshold (right). 
} %
\end{figure}

The results are summarized in \Fig{graphvarconn_perf} (left): each marker therein corresponds 
to a different value of a different criterion, with different criteria corresponding 
to different colors. The position of each marker along the~$x$- and $y$-axes 
corresponds, respectively, to the degree of the resulting interaction graph 
and the total data rate it yields. Ideally, we would like solutions with a low complexity 
(i.e., markers closer to the left-hand side of the plot) and a high data rate 
(i.e., markers closer to the top of the plot). From \Fig{graphvarconn_perf} (left) 
it is clear that red markers, referring to the results obtained under the coverage-based criterion, 
correspond to markedly better trade-offs than blue markers, referring to 
the result achieved under the distance-based criterion. Intuitively, this is equivalent to saying that 
the coverage-based criterion creates {\em better}, i.e.,
 more meaningful edges in the interaction graph, therefore, 
 it yields better performance with the same graph degree (or, equivalently, 
 the same performance with a simpler graph), hence, 
 less overhead and faster convergence of the CRAB algorithm.
 
In terms of convergence, as mentioned in the previous section, CRAB tends to reach the 
state well before the maximum number of iterations, which is set at 50. 
This holds true especially as the connectivity threshold is increased and the 
average node degree is reduced, as shown in \Fig{graphvarconn_perf} (right).

\subsection{Benchmark evaluation}

Next, we move to assessing the performance of our CRAB algorithm against two state-of-the-art benchmarks, 
namely, 
\begin{itemize}
    \item an iterative approach based on avoiding overlapping between beams leveraging 
    conflict-aware weighted bipartite graph matching, presented in~\cite{noi-mobihoc20} 
    and labeled CAWBM in the plots, and 
    \item a clustering-based approach aiming to serve nearby users with the same beam, 
    labeled DBSCAN in the plots.
\end{itemize}
Both benchmark solutions offer a limited flexibility when selecting beam configurations. 
Specifically, CAWBM requires a fixed number of beams, which we set to~$4$, while 
DBSCAN requires a fixed beamwidth, which we set to~$10^\circ$. On the positive side, 
DBSCAN is a distributed algorithm which can be executed dynamically to account 
for mobility in the network, while CAWBM is a static, but centralized,  
algorithm which takes advantage of the comprehensive knowledge about the network to 
better match gNBs with zones.

 For CRAB, we employ the coverage criterion with a threshold 
of~$c_{\rm thr}=0.15$. Further, all messages are initialized to 
random values uniformly drawn between 0 and 1, and we deem that convergence at a single node
is reached when the difference between consecutive outgoing messages is 
lower than $10^{-5}$ times the minimum over all values carried by the most recent messages.

A node will not send out a message that has not changed from the previous iteration. 
A node that has reached convergence will not send out any outgoing messages. 
Finally, a node that has stopped receiving a message from a neighboring node, 
will use the latest received message from that node. 

The network will converge once all the nodes reach the convergent state, and there are no more messages to be exchanged between nodes.  
Again, all algorithms are executed periodically every second, and the total simulation duration 
is 10~s.

\begin{figure*}
\centering
\includegraphics[width=0.32\textwidth]{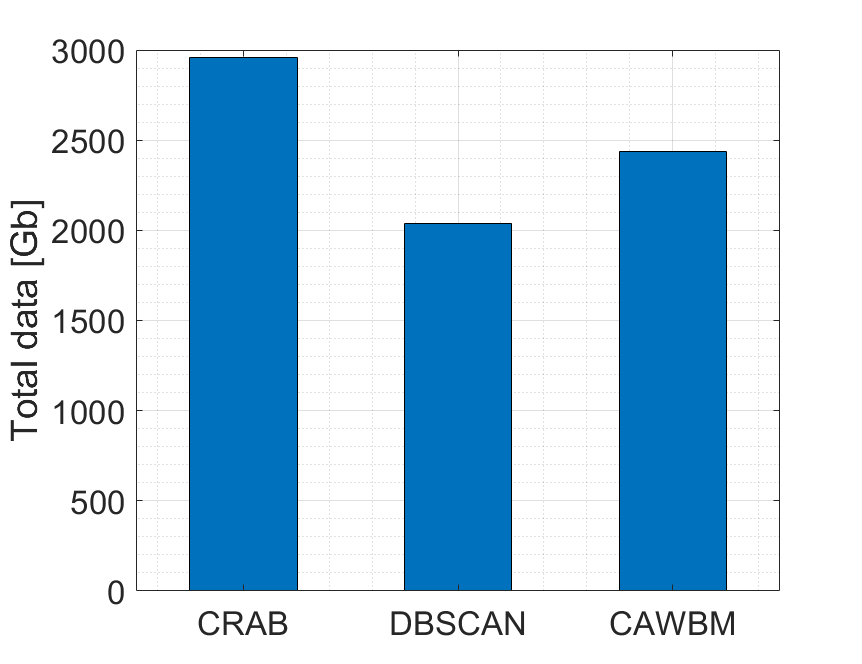}
\includegraphics[width=0.32\textwidth]{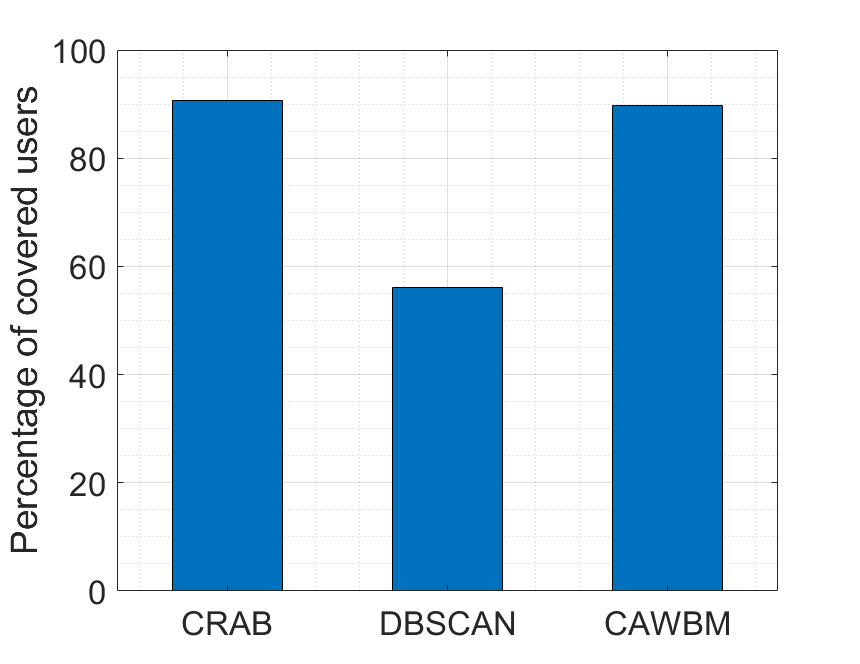}
\includegraphics[width=0.32\textwidth]{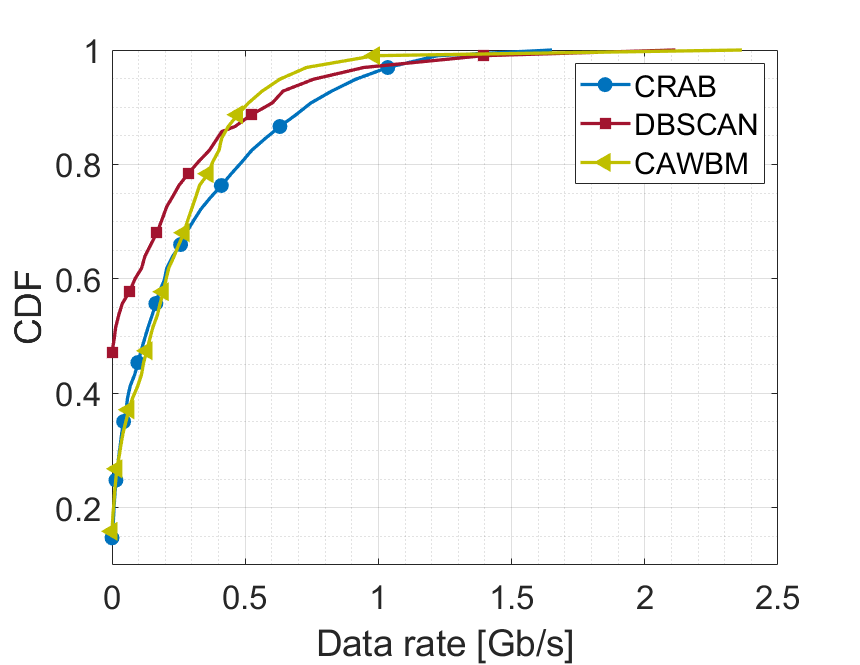}
\caption{\label{fig:bench_overall}Total amount of data downloaded (left); 
percentage of vehicles covered (center); cumulative density function (CDF) of the data rate offered to the vehicles in Gb/s (right).
} %
\end{figure*}

A first aspect we are interested in is the absolute performance, i.e., which of the three 
solutions yields the highest effective data rate.  %
As shown in \Fig{bench_overall}(left), CRAB 
outperforms the benchmark solutions, by delivering in total 45\% more data than DBSCAN, 
and 21\% more than CAWBM.

In order to gather more insight about the performance difference for individual 
vehicles, \Fig{bench_overall}(center) highlights the fraction of vehicles covered by each solution. 
We can immediately notice that one of the reasons for DBSCAN's poor performance is the 
fact that it only manages to serve about 50\% of the vehicles in the network, 
while both CRAB and CAWBM serve more than 80\%. 
Focusing on the users that do get served, \Fig{bench_overall}(right) shows 
the distribution of the data rate obtained by individual vehicles. We can see that 
the difference between CRAB and CAWBM is more clear for the vehicles in the top 
highest percentiles: under CRAB 20\% of the top users can reach data rates over 500 Mb/s, 
while with CAWBM only the top 10\% can reach such data rates. 
In other words, CRAB can serve {\em more} users than DBSCAN, and 
 offer to them a {\em better rate} than CAWBM.
 
 \begin{figure*}
\centering
\includegraphics[width=0.32\textwidth]{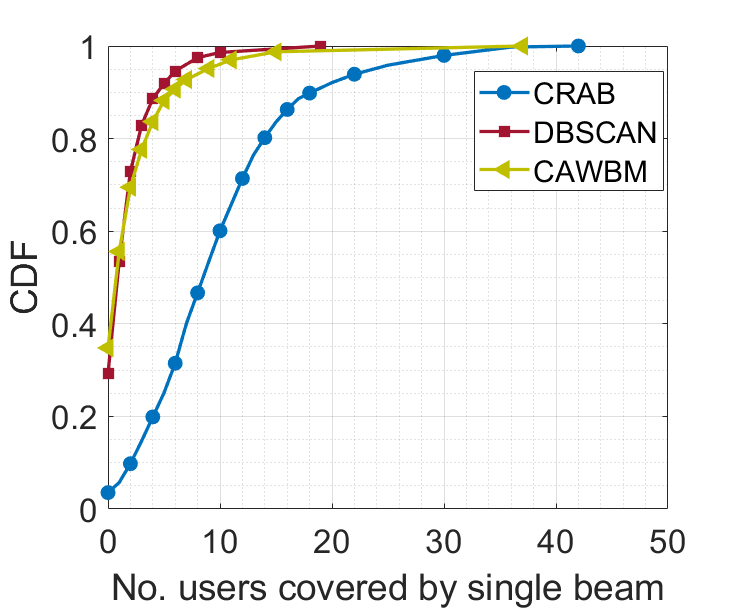}
\includegraphics[width=0.32\textwidth]{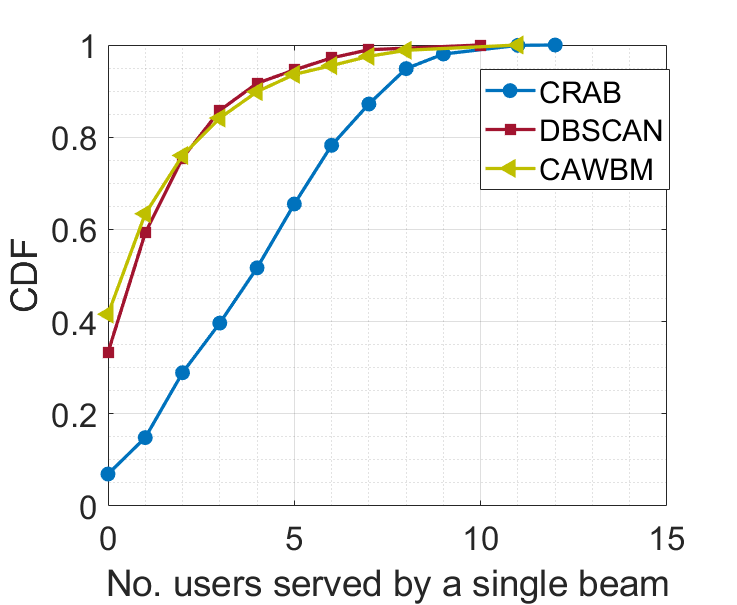}
\includegraphics[width=0.32\textwidth]{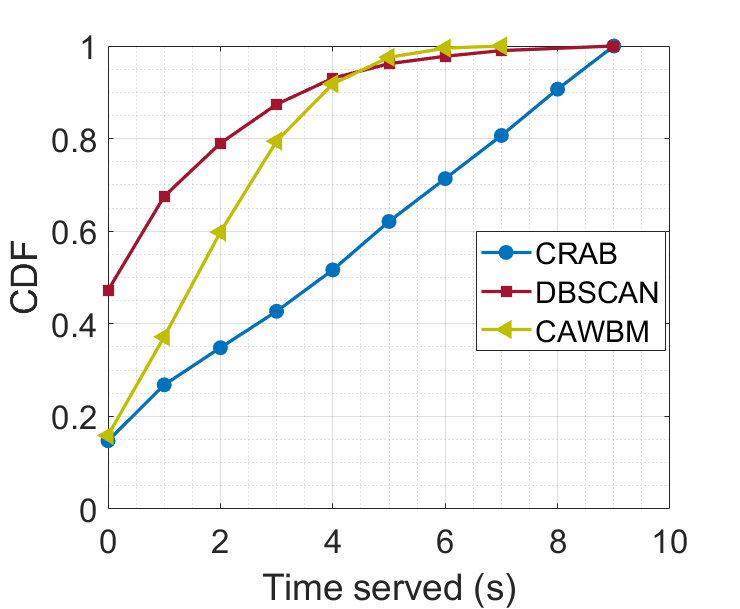}
\caption{\label{fig:bench_users}CDF of number of vehicles covered by individual beams (left); 
CDF of number of vehicles covered by individual beams  (center); CDF of the service duration 
per vehicle in seconds (right). }
\end{figure*}

The high-level reason for such improved performance lies in the better 
coordination between beams and gNBs enabled by the message-passing approach of CRAB. 
\Fig{bench_users}(left) and \Fig{bench_users}(center) highlight two of the main effects 
CRAB's superior beam configuration, namely, that CRAB is able to cover and serve more 
users with each beam than its counterparts, respectively, twice and four times more. 
Also, \Fig{bench_users}(right) shows how CRAB can serve each 
vehicle for a longer time, up to four times the best benchmark. In summary, 
CRAB is able to serve more users, give them a higher  data rate, and cover them for a longer time.

\begin{figure}
\centering
\includegraphics[width=0.45\textwidth]{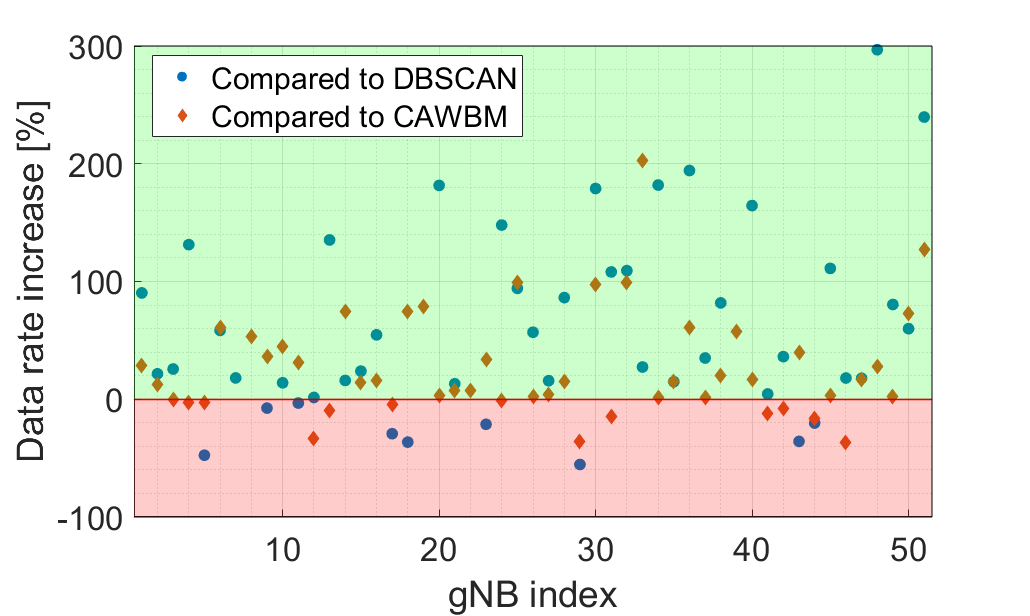} 
\caption{\label{fig:gnb_improvement}Improvement of individual gNBs due to CRAB in terms of  data rate  
compared to DBSCAN (blue circles) and CAWBM (red diamonds).
} %
\end{figure}

As observed by looking at  \Fig{gnb_improvement}, 
while some individual gNBs may see a drop in capacity due to coordination resulting from CRAB, 
more than 80\% of the nodes experience significant gains, often exceeding 
100\%. Specifically, when compared to DBSCAN over 30\% of the gNBs experience improvement 
over 100\%, and 55\% of the nodes experience a gain of 50\% or more. When compared to CAWBM, these numbers are expectedly  lower, respectively, around 10\% of the gNBs experience a data rate increase of 100\% or more, while 30\% of them experience a gain of 50\% or more.

\begin{figure*}
\centering
\includegraphics[width=0.48\textwidth]{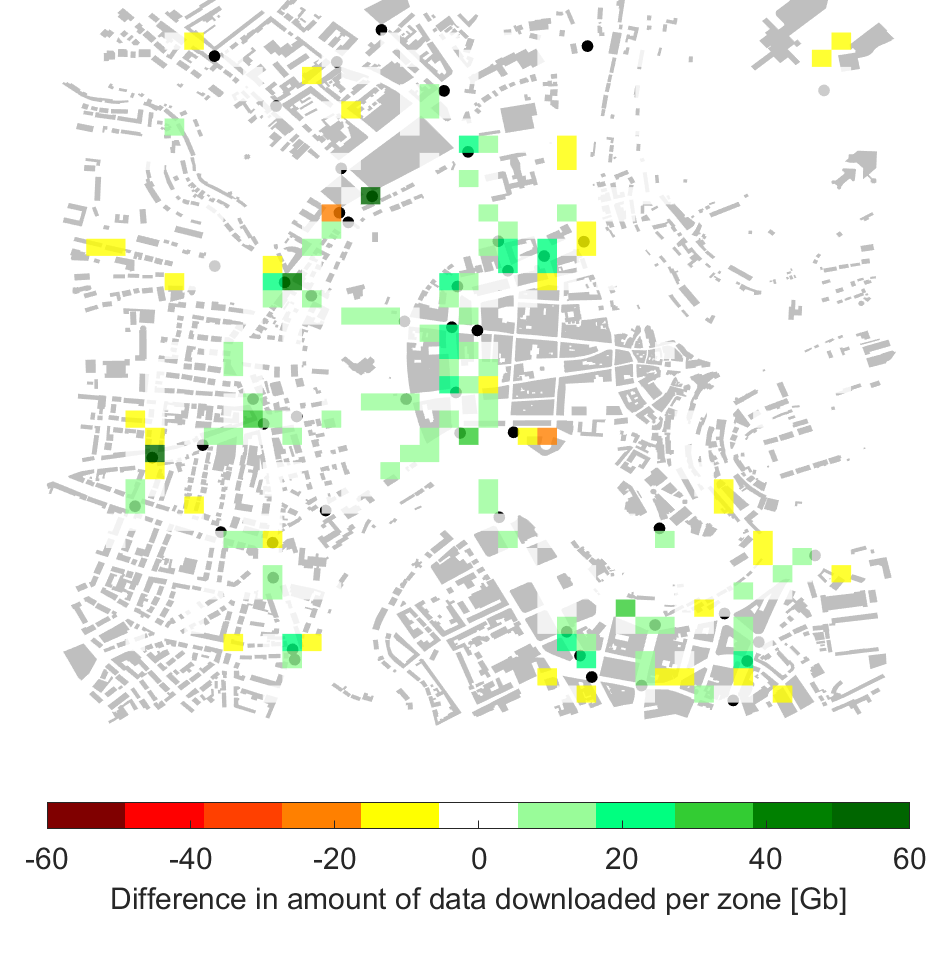}
\includegraphics[width=0.48\textwidth]{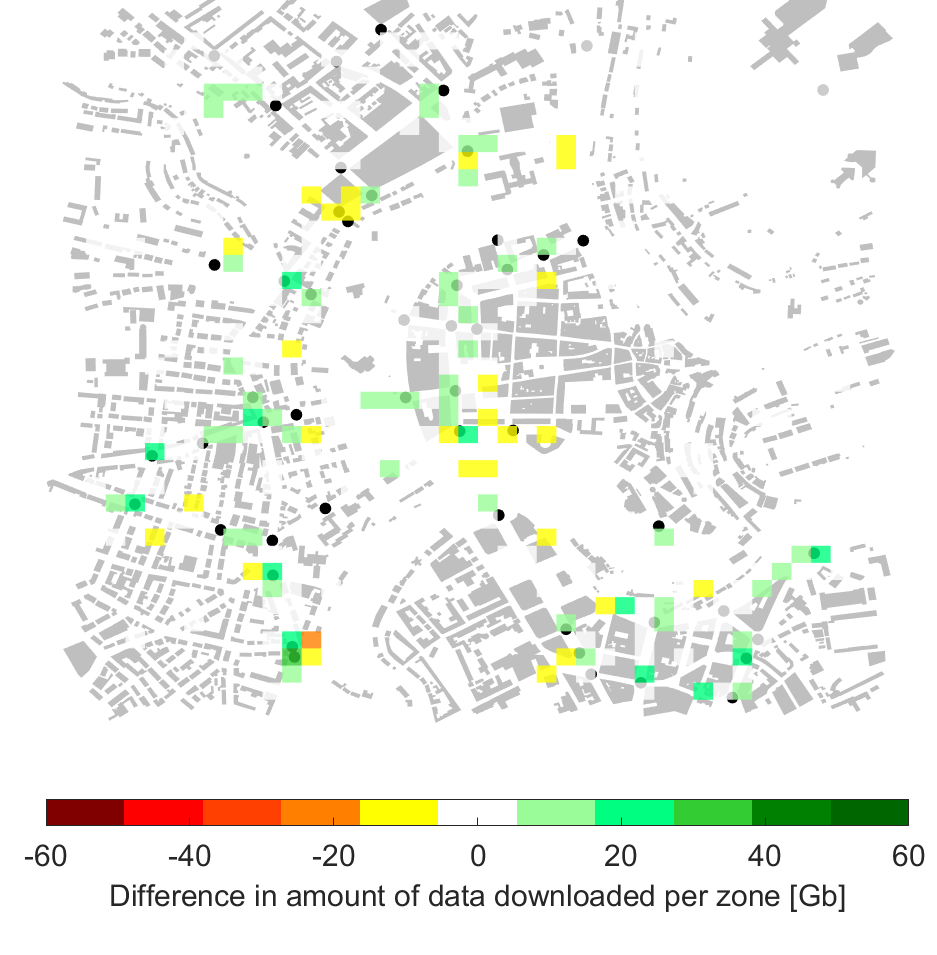}
\caption{\label{fig:bench_improvement}Improvement in absolute terms of amount of data downloaded at 
individual zones: compared to DBSCAN (left); compared to CAWBM (right). }
\end{figure*}

\Fig{bench_improvement} summarizes how the performance obtained by different zones of 
the topology changes as a consequence of moving from DBSCAN to CRAB (left plot) 
and from CAWBM to CRAB (right plot).
Compared to DBSCAN (left plot), CRAB can better serve the areas around the center of 
the city, where both the density of the gNB and vehicles is higher. 
This is due to DBSCAN's fully decentralized nature, whereby gNBs decide the configuration 
of the beams solely on their local knowledge about vehicle mobility. 
When the number of gNBs is relatively high, the configured beams at different 
gNBs may easily target overlapping areas, resulting in inefficient beam usage and 
higher interference. Looking at the right plot and the comparison with CAWBM, 
we can see that the improvement is more evenly distributed across the road topology, 
including the peripheral regions. This is due to CAWBM's centralized approach, 
which works best in dense areas, but may overestimate the interference between 
far-away nodes, thus providing a worse coverage in less dense areas. 
Overall, CRAB emerges as an effective compromise between a fully centralized 
approach like CAWBM and an 
approach like DBSCAN that exploits only local information.

\section{Conclusions}
\label{sec:conclusion}

We  identified mmwave as a promising technology to enhance the
capacity of vehicular networks. However, the performance of mmwave
networks depends on the number, the alignment, and the width of beams
between gNBs and vehicles, and these have to be carefully configured in order to maximize the data rate and 
avoid coverage overlapping among distinct gNBs.

To address this problem, we adopted a randomized approach and formulated 
an optimization problem, providing both a centralized and a decentralized version thereof.
To efficiently find a solution in a distributed manner, we devised an algorithmic framework 
leveraging belief propagation, called CRAB. The proposed framework (i) models the vehicular mmwave 
network as graph, 
capturing the interference that the possible beam configurations at the gNB may generate, 
 (ii) adopts a message passing approach on such a graph, pruning it if necessary, 
and (iii) effectively finds a high-quality solution.  

Our performance evaluation, based on real-world topology and realistic mobility traces, 
shows that CRAB significantly outperforms state-of-the-art alternatives, delivering in total from 21\% up to 
45\% more data, while 
making, on average,  a  single gNB transfers 50\% more data. 
Further, CRAB provides up to 30\% better user coverage and an improved level of fairness in  
data rate  performance across users.

\bibliography{mmwave}
\bibliographystyle{IEEEtran}
\end{document}